\documentclass[a4paper,USenglish]{lipics}

\usepackage{microtype}
\usepackage{comment}
\usepackage{url}

\title{Combinatorial Redundancy Detection
\footnote{Research supported by the Swiss National Science Foundation
  (SNF Project 200021\_150055 / 1)}}
\titlerunning{Combinatorial Redundancy Detection} 

\date{November 30, 2014 (rev KF)}

\author[1]{Komei Fukuda} 
\author[2]{Bernd G\"artner}
\author[3]{May Szedl{\'a}k}
\affil[1]{Department of Mathematics and\\ Institute of Theoretical Computer Science
 ETH Z\"{u}rich, \\ CH-8092 Z\"{u}rich, Switzerland\\
  \texttt{komei.fukuda@math.ethz.ch}}
\affil[2]{Institute of Theoretical Computer Science, ETH Z\"{u}rich\\ CH-8092 Z\"{u}rich, Switzerland\\
  \texttt{gaertner@inf.ethz.ch}}
\affil[3]{Institute of Theoretical Computer Science, ETH Z\"{u}rich\\
  CH-8092 Z\"{u}rich, Switzerland\\
    \texttt{may.szedlak@inf.ethz.ch}}
\authorrunning{K. Fukuda, B. G\"artner and M. Szedl{\'a}k} 

\keywords{system of linear inequalities, redundancy removal, linear programming, output sensitive algorithm, Clarkson's method}

\begin{document}

\maketitle

\begin{abstract}
  The problem of detecting and removing redundant constraints is fundamental in optimization. We focus on the case of linear programs (LPs) in dictionary form, given by $n$ equality constraints in $n+d$ variables, where the variables are constrained to be nonnegative. A variable $x_r$ is called \emph{redundant}, if after removing $x_r \geq 0$ the LP still has the same feasible region. The time needed to solve such an LP is denoted by $LP(n,d)$.

It is easy to see that solving $n+d$ LPs of the above size is sufficient to detect all redundancies. The currently fastest practical method is the one by Clarkson: it solves $n+d$ linear programs, but each of them has at most $s$ variables, where $s$ is the number of nonredundant constraints \cite{c-mosga-94}. 

In the first part we show that knowing all of the finitely many dictionaries of the LP is sufficient for the purpose of redundancy detection. A dictionary is a matrix that can be thought of as an enriched encoding of a vertex in the LP. Moreover --- and this is the combinatorial aspect --- it is enough to know only the signs of the entries, the actual values do not matter. Concretely we show that for any variable $x_r$ one can find a dictionary, such that its sign pattern is either a redundancy or nonredundancy certificate for $x_r$. 

In the second part we show that considering only the sign patterns of the dictionary, there is an output sensitive algorithm of running time $\mathcal{O}(d \cdot (n+d) \cdot s^{d-1} \cdot LP(s,d) + d \cdot s^{d} \cdot LP(n,d))$ to detect all redundancies. In the case where all constraints are in general position, the running time is $\mathcal{O}(s \cdot LP(n,d) + (n+d) \cdot LP(s,d))$, which is essentially the running time of the Clarkson method.  Our algorithm extends naturally 
to a more general setting of arrangements of oriented topological hyperplane arrangements.
\end{abstract}

\newcommand{\R}{\mathbb{R}}

\section{Introduction}
The problem of detecting and removing redundant constraints is fundamental in optimization. Being able to understand redundancies in a model is an important step towards improvements of the model and faster solutions.

In this paper, we focus on redundancies in systems of linear inequalities. We consider
systems of the form 
\begin{equation}\label{eq:system}
\begin{array}{rcl}
x_B &=& b- Ax_N\\
   x_B &\geq& 0 \\
   x_N &\geq&0
\end{array}
\end{equation}
where $B$ and $N$ are disjoint finite sets of variable indices with
$|B|=n$, $|N|=d$, $b\in\R^B$ and $A\in\R^{B\times N}$ are given input vector 
and matrix.  We assume that the system (\ref{eq:system}) has a feasible solution.
Any consistent system of linear equalities and inequalities can be reduced to this form.

A variable $x_r$ is called \emph{redundant} in (\ref{eq:system}) if $x_B = b - Ax_N$ and $x_i\geq 0$ for $i\in B \cup N \setminus\{r\}$ implies $x_r\geq 0$, i.e., if after removing constraint $x_r \geq 0$ from (\ref{eq:system}) the resulting system still has the same feasible region. Testing redundancy of $x_r$ can be done by solving the linear program (LP)
\begin{equation}\label{eq:LP_r}
\begin{array}{lrcl}
\mbox{minimize} & x_r \\ 
\mbox{subject to} & x_B &=& b- Ax_N\\
  & x_i &\geq& 0, \hspace{5mm} \forall i \in B \cup N \setminus \{r\}.   
\end{array}
\end{equation}
Namely, a variable $x_r$ is redundant if and only if the LP has an optimal
solution and the optimal value is nonnegative.

Let $LP(n,d)$ denote the time needed to solve an LP of form (\ref{eq:LP_r}). Throughout the paper, we are working in the real RAM model of computation, where practical algorithms, but no polynomial bounds on $LP(n,d)$ are known. However, our results translate to the standard Turing machine model, where they would involve bounds of the form $LP(n,d,\ell)$, with $\ell$ being the bit size of the input. In this case, $LP(n,d,\ell)$ can be polynomially bounded. The notation $LP(n,d)$ abstracts from the concrete representation of the LP, and also from the algorithm being used; as a consequence, we can also apply it in the context of LPs given by the signs of their dictionaries.

By solving $n+d$ linear programs, $\mathcal{O}((n+d)\cdot LP(n,d))$ time is enough to detect all redundant variables in the real RAM model, but it is natural to ask whether there is a faster method. The currently fastest practical method is the one by Clarkson with running time $\mathcal{O}((n+d)\cdot LP(s,d) + s \cdot n \cdot d)$~\cite{c-mosga-94}. This method also solves $n+d$ linear programs, but each of them has at most $s$ variables, where $s$ is the number of nonredundant variables. Hence, if $s\ll n$, this \emph{output-sensitive} algorithm is a major improvement.

A related (dual) problem is the one of finding the {extreme points} among a set $P$ of $n$ points in $\R^d$. A point $p\in P$ is {\em extreme\/} in $P$, if $p$ is not contained in the convex hull of $P\setminus\{p\}$. It is not hard to see that this problem is a special case of redundancy detection in linear systems.

Specialized (and output-sensitive) algorithms for the extreme points problem exist ~\cite{oss-eephd-95,MR1649794}, but they are essentially following the ideas of Clarkson's algorithm~\cite{c-mosga-94}. For fixed $d$, Chan uses elaborate data structures from computational geometry to obtain a slight improvement over Clarkson's method~\cite{Chan}.

In this paper, we study the \emph{combinatorial} aspects of redundancy detection in linear systems. The basic questions are: What kind of information about the linear system do we need in order to detect all redundant variables? With this restricted set of information, how fast can we detect all of them? Our motivation is to explore and understand the boundary between geometry and combinatorics with respect to redundancy. For example, Clarkson's method~\cite{c-mosga-94} uses \emph{ray shooting}, an intrinsically geometric procedure; similarly, the dual extreme points algorithms~\cite{oss-eephd-95,MR1649794} use scalar products. In a purely combinatorial setting, neither ray shooting nor scalar products are well-defined notions, so it is natural to ask whether we can do without them.

Our approach is very similar to the combinatorial viewpoint of linear programming pioneered by Matou\v{s}ek, Sharir and Welzl~\cite{msw-sblp-92} in form of the concept of \emph{LP-type problems}. The question they ask is: how quickly can we \emph{optimize}, given only combinatorial information? As we consider redundancy detection and removal as important towards efficient optimization, it is very natural to extend the combinatorial viewpoint to also include the question of redundancy. The results that we obtain are first steps and leave ample space for improvement. An immediate theoretical benefit is that we can handle redundancy detection in structures that are more general than systems of linear inequalities; most notably, our results naturally extend to the realm of \emph{oriented matroids}~\cite{blswz-om-93}. 

\paragraph*{Statement of Results.}
The first point that we will make, is that for the purpose of redundancy testing, it is sufficient to know all the finitely many dictionaries associated with the system of inequalities (\ref{eq:system}). 
A dicitonary can be thought of as an encoding of the associated arrangements of hyperplanes (see Section \ref{sec:basics}). 
Moreover, we show that it is sufficient to know only the \emph{signed} dictionaries, i.e., the \emph{signs} of the dictionary entries. Their actual numerical values do not matter.

In Theorem \ref{thm:redundant}, we give a characterization of such a redundancy certificate. 
More precisely, we show that, for every redundant variable $x_r$ there exists at least one signed dictionary such that its sign pattern is a redundancy certificate of $x_r$. 
Similarly, as shown in Theorem \ref{thm_nonred}, for every nonredundant variable there exists a nonredundancy certificate. Such a single certificate can be detected in time $LP(n,d)$ (see Section \ref{sec_criss}). 
The number of dictionaries needed to detect \emph{all} redundancies depends on the LP and can vary between constant and linear in $n+d$ (see Appendix \ref{sec:examples}).

In a second part, we present a Clarkson-type, output-sensitive algorithm that detects all redundancies in running time $\mathcal{O}(d\cdot  (n+d) \cdot s^{d-1} LP(s,d) + d \cdot s^d \cdot LP(n,d))$ (Theorem \ref{thm_algo}). Under some general position assumptions the running time can be improved to $\mathcal{O}((n+d) \cdot LP(s,d) + s \cdot LP(n,d))$, which is basically the running time of Clarkson's algorithm. In these bounds, $LP(d,n)$ denotes the time to solve an LP to which we have access only through signed dictionaries. As in the real RAM model, no polynomial
bounds are known, but algorithms that are fast in practice exist.

In general our algorithm's running time is worse than Clarkson's, but it only requires the combinatorial information of the system and not its actual numerical values.  If the feasible region is not full dimensional (i.e.\ not of dimension $d$), then a redundant constraint may become nonredundant after the removal of some other redundant constraints. To avoid these dependencies of the redundant constraints we assume full dimensionality of the feasible region. Because of our purely combinatorial characterizations of redundancy and nonredundancy, our algorithm works in the combinatorial setting of oriented matroids \cite{blswz-om-93}, and can be applied to remove redundancies from oriented topological hyperplane arrangements.


\section{Basics}\label{sec:basics}

Before discussing redundancy removal and combinatorial aspects in linear programs, we fix the basic notation on linear programming, --- such as dictionaries and pivots steps --- and review finite pivot algorithms. (For further details and  proofs see e.g.\ \cite{chvatal-LP, fukuda-optimization}.)

\subsection{LP in Dictionary Form}

Throughout, if not stated otherwise, we always consider \emph{linear programs} (LPs) of the form
\begin{equation}\label{LP1}
\begin{array}{lrcl}
\mbox{minimize} & c^Tx_N \\ 
\mbox{subject to} & x_B &=& b- Ax_N\\
  & x_i &\geq& 0, \hspace{5mm} \forall i \in E:=B \cup N,   
\end{array}
\end{equation}
where as introduced in (\ref{eq:system}), 
$B$ and $N$ are disjoint finite sets of variable indices with
$|B|=n$, $|N|=d$, $b\in\R^B$ and $A\in\R^{B\times N}$ are given input vector 
and matrix. An LP of this form is called LP in \emph{dictionary form} and its \emph{size} is $n \times d$. The set $B$ is called a (initial) \emph{basis}, $N$ a (initial) \emph{nonbasis} and $c^Tx_N$ the \emph{objective function}. 

The \emph{feasible region} of the LP is defined as the set of $x \in \R^E$ that satisfy all constraints, i.e., the set
$\{x \in \R^E | x_B = b - Ax_N, x_i \geq 0, \forall i \in E\}$. A feasible solution $\overline{x}$ is called \emph{optimal} if for every feasible solution $x$, $c^T\overline{x} \leq c^T x$. The LP is called \emph{unbounded} if for every $k \in \R$, there exists a feasible solution $x$, such that $c^Tx \leq k$. If there exists no feasible solution, the LP is called \emph{infeasible}.

The \emph{dictionary} $D(B) \in \R^{B\cup\{f\} \times N\cup\{g\}}$ of an LP (\ref{LP1}) w.r.t.\ a basis $B$ is defined as
\begin{align*}  
D : = D(B) & = 
\left[
 \begin{array}{rr}
   0  &  c^T \\
   b  &  -A   \\
  \end{array}
\right],
\end{align*}
where $f$ is the index of the first row and $g$ is the index of the first column.
For each $i \in B \cup \{f\}$ and $j \in N \cup \{g\}$, we denote by $d_{ij}$ its $(i,j)$ entry, by $D_{i.}$ the row indexed by $i$, and by $D_{.j}$ the column indexed by $j$. 

Hence by setting $x_f: = c^Tx_N$, we can rewrite ($\ref{LP1}$) as 
\begin{equation}
\begin{array}{lrcl}
\mbox{minimize} & x_f\\ 
\mbox{subject to} & x_{B \cup \{f\}} &=& Dx_{N \cup \{g\}}\\
  & x_i &\geq& 0, \hspace{5mm} \forall i \in E:=B \cup N \\
  & x_g &=& 1.
\end{array}
\end{equation}
Whenever we do not care about the objective function, 
we may set $c=0$, and with abuse of notation, set $D = [b, -A]$.

The {\em basic solution} w.r.t.\ $B$ is the unique solution $\overline{x}$ to $x_{B \cup \{f\}} = Dx_{N \cup \{g\}}$ such that $\overline{x}_g = 1$, $\overline{x}_N = 0$ and hence $\overline{x}_{B \cup \{f\}} = D_{.g}$.
 
It is useful to define the following four different types of dictionaries (and bases) as shown in the figure below, where "$+$" denotes positivity, "$\oplus$" nonnegativity and similarly "$-$" negativity and "$\ominus$" nonpositivity.

A dictionary $D$
(or the associated basis $B$) is called {\em feasible\/} if
$d_{ig} \ge 0$ for all $i \in B$.
A dictionary $D$
(or the associated basis $B$) is called {\em optimal\/} if
$d_{ig} \ge 0$, $d_{fj} \ge 0$ for all $i \in B, j \in N$.
A dictionary $D$
(or the associated basis $B$) is called {\em inconsistent\/} if
there exists $r \in B$ such that $d_{rg} < 0$ and
$d_{rj} \le 0$ for all $j \in N$.
A dictionary $D$
(or the associated basis $B$) is called {\em dual inconsistent\/} if
there exists $s \in N$ such that $d_{fs} < 0$ and
$d_{is} \ge 0$ for all $i \in B$.

\begin{figure*} [ht]    

\begin{centering}
\begin{tabular}{r|c|c|l}
  \multicolumn{1}{r}{ } & \multicolumn{1}{c}{$g$} 
& \multicolumn{2}{c}{ } \\  \cline{2-3}
 $f$  &   &        &  \\ \cline{2-3}
      & $\oplus$   &                &    \\
      & $\vdots$   &  \hspace{20mm} &    \\
      & $\oplus$   &                &    \\ \cline{2-3}
  \multicolumn{4}{c}{feasible}  \\
\multicolumn{4}{r}{ }\\
\multicolumn{4}{r}{ }\\
  \multicolumn{1}{r}{ } & \multicolumn{1}{c}{$g$} 
& \multicolumn{1}{c}{ } &  \\ \cline{2-3}
  $f$ &     &                &    \\ \cline{2-3}
      &     &   \hspace{20mm}  & \\
 $\exists r$  
      & $-$ & $\ominus \quad \cdots \quad \ominus\!$ &  \\ 
      &     &                &     \\  \cline{2-3}
  \multicolumn{4}{c}{inconsistent} \\
\end{tabular} \hspace{10mm}
\begin{tabular}{r|c|c|l}
  \multicolumn{1}{r}{ } & \multicolumn{1}{c}{$g$} 
& \multicolumn{2}{c}{ } \\  \cline{2-3}
 $f$  &    & $\oplus \quad \cdots \quad \oplus\!$ &  \\ \cline{2-3}
      &  $\oplus$  &                &    \\
      &  $\vdots$  &  \hspace{20mm} &    \\
      &  $\oplus$  &                &    \\ \cline{2-3}
  \multicolumn{4}{c}{optimal}  \\
\multicolumn{4}{r}{ }\\
\multicolumn{4}{r}{ }\\
  \multicolumn{1}{r}{ } & \multicolumn{1}{c}{$g$} 
&   \multicolumn{1}{c}{$\exists  s$}  &  \\ \cline{2-3}
  $f$ &     &  \hspace{8mm} $-$  \hspace{8mm}  &    \\ \cline{2-3}
      &     &    $\oplus$  & \\
      &     &    $\vdots$  & \\ 
      &     &    $\oplus$  & \\  \cline{2-3}
  \multicolumn{4}{c}{dual inconsistent} \\
\end{tabular}  

\end{centering}
\end{figure*}

The following proposition, follows from standard calculations.
\begin{Proposition}
\label{prop_duality}
For any LP in dictionary form the following statements hold.
\begin{enumerate}
\item
If the dictionary is feasible then the associated basic solution is feasible.
\item
If the dictionary is optimal, then the associated basic solution is optimal.
\item
If the dictionary is inconsistent, then the LP is infeasible.
\item
If the dictionary is dual inconsistent, then the dual LP is infeasible. If in addition the LP is feasible, then the LP is unbounded.
\end{enumerate}
\end{Proposition}

\subsection{Pivot Operations}
We now show how to transform the dictionary of an LP into a modified dictionary using elementary matrix operation, preserving the equivalence of the associated linear system. This operation is called a \emph{pivot operation}.

Let $r \in B$, $s \in N$ and $d_{rs} \neq 0$. Then it is easy to see that one can transform  $x_{B \cup \{f\}} = Dx_{N \cup \{g\}}$ to an equivalent system (i.e., with the same solution set)  :
\[x_{B' \cup \{f\}} = D'x_{N' \cup \{g\}},\] 
where $B' = B \setminus \{r\} \cup \{s\}$ ($N' = N \setminus \{s\} \cup \{r\}$, respectively)
is a new (non)basis and
\begin{gather} \label{LPT:dicpiv3}
d'_{ij}=  
\begin{cases}
   \frac{1}{d_{rs}}  & \text{if } i=s \text{ and } j=r  \\
   - \frac{d_{rj}}{d_{rs}}  & \text{if } i=s \text{ and } j\neq r  \\  
   \frac{d_{is}}{d_{rs}}   & \text{if } i\neq s \text{ and } j= r  \\  
   d_{ij}- \frac{d_{is} \cdot d_{rj}}{d_{rs}}   & \text{if } i\neq s \text{ and } j\neq r \\  
\end{cases}
(i\in B'\cup \{f\} \text{ and }  j\in N' \cup \{g\}).
\end{gather}

We call a dictionary \emph{terminal} if it is optimal, inconsistent or dual inconsistent. There are several finite pivot algorithms such as the simplex and the criss-cross method that transform any dictionary into one of the terminal dictionaries \cite{terlaky-criss, ft-ccmfv-97, dantzig-simplex}. This will be discussed further in Section \ref{sec_criss}.                                                                                                                                                                                                                                                                                                                                                                                                                                                                              
\section{Combinatorial Redundancy}

Consider an LP in dictionary form as given in (\ref{LP1}).
Then $x_r \geq 0$ is \emph{redundant}, if the removal of the constraint does not change the feasible solution set, i.e., if
\begin{equation}\label{LPred2}
\begin{array}{lrcl}
\mbox{minimize} & c^Tx_N \\ 
\mbox{subject to} & x_B &=& b- Ax_N\\
  & x_i &\geq& 0, \hspace{5mm} \forall i \in E \setminus \{r\},   
\end{array}
\end{equation}
has the same feasible solution set as (\ref{LP1}). Then the variable $x_r$ and the index $r$ are called \emph{redundant}. 

If the constraint $x_r \geq 0$ is not redundant it is called \emph{nonredundant}, in that case the variable $x_r$ and the index $r$ are called \emph{nonredundant}.

It is not hard to see that solving $n+d$ LPs of the same size as (\ref{LPred2}) suffices to find all redundancies. 
Hence running time $\mathcal{O}((n+d) \cdot LP(n,d))$ suffices to find all redundancies, where $LP(n,d)$ is the time needed to solve an LP of size $n \times d$. 
Clarkson showed that it is possible to find \emph{all} redundancies in time 
$\mathcal{O}((n+d)\cdot LP(s,d) + s\cdot n\cdot d)$, where $s$ is the number of nonredundant variables \cite{c-mosga-94}. In case where $s \ll n$ this is a major improvement. To be able to execute Clarkson's algorithm, one needs to assume full dimensionality and an interior point of the feasible solution set. In the LP setting this can be done by some preprocessing, including solving a few ($O(d)$) LPs \cite{fukuda-polyhedral}. 

In the following we focus on the combinatorial aspect of redundancy removal. 
We give a combinatorial way,  the \emph{dictionary oracle}, to encode LPs in dictionary form, where we are basically only given the signs of the entries of the dictionaries. In Section \ref{sec:certificates} we will show how the signs suffice to find all redundant and nonredundant constraints of an LP in dictionary form. 


Consider an LP of form (\ref{LP1}).
For any given basis $B$, the \emph{dictionary oracle} returns a matrix
\[D^{\sigma} = D^{\sigma}(B) \in \{+,-,0\}^{B\times N\cup\{g\}}, \text{ with }
d^{\sigma}_{ij} = sign({d_{ij}}), \forall i \in B, j \in N\cup\{g\}.\]
Namely, for basis $B$, the oracle simply returns the matrix containing the signs of $D(B)$, without the entries of the objective row $f$. For combinatorial redundancy detection the objective function is not needed since redundancy of a constraint only depends on the given set of linear inequalities.

\section{Certificates}
\label{sec:certificates}
We show that the dictionary oracle is enough to detect all redundancies and nonredundancies of the variables in $E$. More precisely for every $r \in E$, there exists a basis $B$ such that $D^{\sigma}(B)$ is either a redundancy or nonredundancy certificate for $x_r$. We give a full characterization of the certificates in Theorem \ref{thm:redundant} and Theorem \ref{thm_nonred}. The number of dictionaries needed to have \emph{all} certificates depend on the LP. See the Appendix \ref{sec:examples} for examples where constantly many suffice and where linearly many are needed.

For convenience throughout we make the following assumptions, which can be satisfied with 
simple preprocessing.

\begin{enumerate}
\item
\label{ass_nonempty}
The feasible region of (\ref{LP1}) is full dimensional (and hence nonempty).
\item
\label{ass_parallel}
There is no $j \in N$ such that $d_{ij} =0$ for all $i \in B$.
\end{enumerate}

In Section \ref{sec_criss} we will see that both the criss-cross and the simplex method can be used on the dictionary oracle for certain objective functions. Testing whether the feasible solution set is empty can hence be done by solving one linear program in the oracle setting. As mentioned in the introduction the full-dimensionality assumption is made to avoid dependencies between the redundant constraints. This can be achieved by some preprocessing on the LP, including solving a few ($O(d)$) LPs \cite{fukuda-polyhedral}.

It is easy to see that if there exists a column $j$ such that $d_{ij} =0$ for all $i \in B$, then $x_j$ is nonredundant and we can simply remove the column.

\subsection{A Certificate for Redundancy in the Dictionary Oracle}
We say a that basis $B$ is \emph{$r$-redundant} if $r \in B$ and $D^{\sigma}_{r.} \geq 0$ i.e.\ if $D^{\sigma}(B)$ is as given in the figure below. 

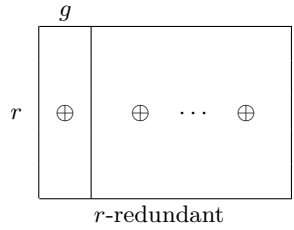
\begin{figure} [ht]    

\begin{center}
\begin{tabular}{r|c|c|l}
  \multicolumn{1}{r}{ } & \multicolumn{1}{c}{$g$} 
& \multicolumn{1}{c}{ } &  \\ \cline{2-3}
      &     &   \hspace{22mm}  & \\
      &     &    & \\
 $r$  & $\oplus$ & $\oplus \quad \cdots \quad \oplus\!$ &  \\ 
      &     &    & \\
     &     &                &     \\  \cline{2-3}
  \multicolumn{4}{c}{$r$-redundant} \\
\end{tabular}
\end{center}
\end{figure}

 Since the $r$-th row of the dictionary represents $x_r = d_{rg} + \sum_{j \in N} d_{rj}x_j$, $x_r \geq 0$ is satisfied as long as $x_j \geq 0$ for all $j \in N$. Hence $x_r \geq 0$ is redundant for (\ref{LP1}).  

\begin{Theorem}[\textbf{Redundancy Certificate}] \label{thm:redundant}
An inequality $x_r \ge 0$ is redundant for the system (\ref{LP1})
if and only if there exists an $r$-redundant basis.
\end{Theorem}
\begin{proof}
We only have to show the ``only if'' part.  

Suppose $x_r \ge 0$ is redundant for the system (\ref{LP1}).  We will show that there exists an $r$-redundant basis.

Consider the LP minimizing the variable $x_r$ subject to the system (\ref{LP1})
without the constraint $x_r \ge 0$.   Since $x_r \ge 0$ is redundant for the system (\ref{LP1}), the LP is bounded. 
By assumption \ref{ass_nonempty} and the fact that every finite pivot algorithm terminates in a terminal dictionary the LP has an optimal dictionary.

If the initial basis contains $r$, then we can consider the row associated with $r$
as the objective row.  Apply any finite pivot algorithm to the LP.  Otherwise,
$r$ is nonbasic.  By assumption \ref{ass_parallel}, one can pivot on the $r$-th column to make
$r$ a basic index. This reduces the case to the first case.

Let's consider an optimal basis and optimal dictionary for the LP where $x_r$ is the objective function.
Since it is optimal, all entries $d_{rj}$ for $j \in N$ are nonnegative. Furthermore, $d_{rg}$ is nonnegative as otherwise we would have found a solution that satisfies all constraints except $x_r \geq 0$, implying nonredundancy of $x_r$. 
\end{proof}

From the proof of Theorem \ref{thm:redundant} the following strengthening of Theorem \ref{thm:redundant} immediately follows.

\begin{Corollary} \label{cor_redundant}
An inequality $x_r \ge 0$ is redundant for the system (\ref{LP1})
if and only if there exists a \emph{feasible} $r$-redundant basis.
\end{Corollary}

\subsection{A Certificate for Nonredundancy in the Dictionary Oracle}
\label{sec_certificate}
Similarly as in the redundancy case, we introduce a certificate for nonredundancy using the dictionary oracle.
A basis $B$ is called \emph{$r$-nonredundant} if $B$ is feasible, $r \in N$ and $d_{tg} = 0$ implies $d_{tr} \leq 0$ for all $t \in B$ i.e.\ $D^{\sigma}(B)$ is of the following form.

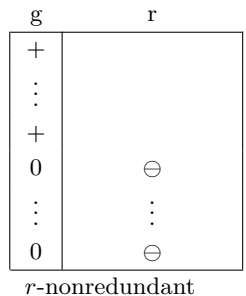
\begin{center}
\begin{tabular}{c |c| c c c c c| }
\multicolumn{1}{r}{}
 &  \multicolumn{1}{c}{g}
 &  \multicolumn{1}{c}{}
 & \multicolumn{1}{c}{}
  & \multicolumn{1}{c}{r}
  & \multicolumn{1}{c}{} 
  & \multicolumn{1}{c}{} \\
\cline{2-7}

& + & {}&{} &{} &{} &{} \\

& $\vdots$& {}&{} &{} &{} &{} \\

& + & {}&{} &{} &{} &{} \\

&$0$ & {} & {} & $\ominus$ & {} & {}\\
&$\vdots$ & {} & {} & $\vdots$ & {} & {}\\
&$0$ & {} & {} & $\ominus$ & {} & {}\\

\cline{2-7} 
\multicolumn{7}{c}{$r$-nonredundant} \\
\end{tabular}
\end{center}

\begin{Theorem}[\textbf{Nonredundancy Certificate}]
\label{thm_nonred}
An inequality $x_r \geq 0$ is nonredundant for the system (\ref{LP1}) if and only if there exists an $r$-nonredundant basis.
\end{Theorem}

Before proving the theorem, we observe the following.
\begin{enumerate}
\item
Unlike in the redundancy certificate an $r$-nonredundant basis needs to be feasible. To verify the correctness of a nonredundancy certificate we need to check between $n$ and $2n$ entries, which is typically much larger than the $d+1$ entries we need for the redundant case. 

\item
If the $g$-column of a feasible basis does not contain any zeros, then all nonbasic variables are nonredundant. 
In general when $x_r \geq 0$ is nonredundant, not necessarily every feasible basis $B$ with $r \in N$ is $r$-nonredundant. Consider the system:
\begin{align}
& x_3 = x_1 + x_2 \nonumber\\
& x_1, x_2, x_3 \geq 0. \nonumber
\end{align}

Then the basis $\{3\}$ is not a certificate of nonredundancy of $x_1$, as $d^{\sigma}_{31} = +$ in the associated dictionary. On the other hand, the basis $\{2\}$ is $1$-nonredundant:
\begin{center}
\begin{tabular}{ c|c|c c| }
\multicolumn{1}{r}{}
 &  \multicolumn{1}{c}{g}
 & \multicolumn{1}{c}{1}
  & \multicolumn{1}{c}{2} \\
\cline{2-4}
3 & 0 & $+$ & $+$\\
\cline{2-4}
\end{tabular}
\hspace{8mm}
\begin{tabular}{ c|c|c c| }
\multicolumn{1}{r}{}
 &  \multicolumn{1}{c}{g}
 & \multicolumn{1}{c}{1}
  & \multicolumn{1}{c}{3} \\
\cline{2-4}
2 & 0 & $-$ & $+$\\
\cline{2-4}
\end{tabular}
\end{center}

\end{enumerate}

\begin{proof}[Proof of Theorem \ref{thm_nonred}]
Let $(LP)$ be of form (\ref{LP1}) and suppose that $x_r \geq 0$ is nonredundant. Consider $(LP)$ without this constraint i.e.\
\begin{equation}
\begin{array}{lrcl}
\mbox{minimize} & x_r \\ 
\mbox{subject to} & x_B &=& b- Ax_N\\
  & x_i &\geq& 0, \hspace{5mm} \forall i \in B \cup N \setminus \{r\}.   
\end{array}
\end{equation}
Then this LP either has optimal solution $-c < 0$ or is unbounded. In the first case choose $0 < \epsilon < c$, in the latter $0 < \epsilon < \infty$ and consider the following perturbed version of $(LP)$, denoted $(LP^\epsilon)$.
\begin{equation} \label{LP_epsilon}
\begin{array}{lrcl}
\mbox{minimize} & x_r \\ 
\mbox{subject to} & x_B &=& b- Ax_N\\
  & x_i &\geq& 0, \hspace{5mm} \forall i \in B \cup N \setminus \{r\} \\
  & x_r &\geq& -\epsilon.   
\end{array}
\end{equation}
Note that this LP can easily be transformed to an LP of form (\ref{LP1}) by the straight forward variable substitution $x'_r=x_r+\epsilon$.

Clearly, LP (\ref{LP_epsilon}) has optimal solution $x_r = -\epsilon$ ($x'_r=0$) and there exists an optimal dictionary where $r$ is a nonbasic variable. 
This follows because if $r$ is basic in an optimal dictionary, then $x_r = -\epsilon$, by choice of $\epsilon$ and after any pivot step in the $r$-th row, the updated basis corresponds to the same basic feasible solution.

Therefore we know there exists a feasible basis $B$ of $(LP^\epsilon)$ with $r \in N$ that minimizes $x_r$. We show that if we choose $\epsilon$ small enough, $B$ is $r$-nonredundant in $(LP)$. Let $B_1,B_2,\dots,B_m$ be the set of all bases (feasible and infeasible) of $(LP)$, that have $r$ as a nonbasic variable. Choose $\epsilon>0$ such that

\[\epsilon < \min_{{i=1,2,\dots,m} \atop {t \in B_i: d_{tg}, d_{tr} < 0}}\frac{d_{tg}}{d_{tr}}.\]

If the RHS is undefined, we choose any $\epsilon < \infty$.

Geometrically this means that there exists no $t \in B_i$ such that $x_t \geq 0$ is violated in the basic solution corresponding to $B_i$ in $(LP)$, but satisfied in the corresponding basic solution in $(LP^{\epsilon})$.

Let $D$ and $D^{\epsilon}$ be the dictionaries w.r.t. $B$ in $(LP)$ and $(LP^{\epsilon})$ respectively.

$D$ and $D^\epsilon$ only differ in their entries of column $g$, where
\[ d_{tg}^\epsilon = d_{tg} - \epsilon \cdot d_{tr}, \forall t \in B.\]

We need to show that $d_{tg} \geq 0$ for all $t \in B$. If $d_{tr} \geq 0$, then this is clear. In the case where $d_{tr} < 0$ it follows that $\epsilon \geq \frac{d_{tg}}{d_{tr}}$ and hence $d_{tg} \geq 0$ by choice of $\epsilon$.


For the other direction let $B$ be $r$-nonredundant and $D$ and $D^\epsilon$ the corresponding dictionaries in $(LP)$ and $(LP^\epsilon)$, respectively. Choose $\epsilon>0$ such that
\[\epsilon \leq \min_{{t \in B} \atop {d_{tg} >0, d_{tr} > 0}}\frac{d_{tg}}{d_{tr}}.\]
If the RHS is undefined, we choose any $\epsilon < \infty$.

We claim that for such an $\epsilon$, $B$ is still feasible for $(LP^\epsilon)$ and hence $x_r \geq 0$ is nonredundant. Again the two dictionaries only differ in row $g$, where
\[ d_{tg}^\epsilon = d_{tg} - \epsilon \cdot d_{tr}, \forall t \in B.\]

In the case where $d_{tg} =0$, it follows that $d_{tg}^{\epsilon} \geq 0$ by $r$-nonredundancy. If $d_{tg} > 0$, then
\[d_{tg}^{\epsilon} = d_{tg} - \epsilon \cdot d_{tr} \geq d_{tg} - \min_{{t' \in B} \atop {d_{t'g} >0, d_{t'r} > 0}} \frac{d_{t'g}}{d_{t'r}} \cdot {d_{tr}} \geq 0.\]

\end{proof}

\subsection{Finite Pivot Algorithms for Certificates}
\label{sec_criss}
In this section we discuss how to design finite pivot algorithms for the dictionary oracle model. 
Both the criss-cross method and the simplex method can be used for the dictionary oracle to find redundancy and nonredundancy certificates.  
A finite pivot algorithm chooses in every step a pivot according to some given rule and terminates in an optimal, inconsistent or dual inconsistent basis in a finite number of steps. 
Note that both the criss-cross method and the simplex method may not be polynomial in the worst case, but are known to be fast in practice \cite{klee-minty-simplex, roos-crisscross}. Furthermore there exits no known polynomial algorithm to solve an LP given by the dictionary oracle. Fukuda conjectured that the randomized criss-cross method is an expected polynomial time algorithm \cite{f-woanfr-11}.

By the proof of Theorem \ref{thm:redundant}, in order to find a redundancy certificate in (\ref{LP1}) it is enough to solve (\ref{LP1}) with objective function $x_r$. Similarly by the proof of Theorem \ref{thm_nonred}, for a nonredundancy certificate it is enough to solve the $\epsilon$-perturbed version (\ref{LP_epsilon}).

For the criss-cross method, the pivot rule is solely dependent on the signs of the dictionary entries and not its actual values \cite{terlaky-criss, ft-criss, chvatal-LP}. 
Standard calculations show that the signs in the $\epsilon$-perturbed dictionary (for $\epsilon > 0$ small enough) are completely determined by the signs of the original dictionary. 
We recall that the dictionary oracle does not output the objective row, but since we minimize in direction of $x_r$ the signs of the objective row are completely determined. (If $r$ is basic then the objective row has the same entries as the $r$-th row and if $r$ nonbasic then $d^{\sigma}_{fr} = +$ and all other entries of the objective row are zero.)  Therefore the dictionary oracle is enough to decide on the pivot steps of the criss-cross method. 

For the simplex method with the smallest index rule, we are given a feasible basis and the nonbasic variable of the pivot element is chosen by its sign only \cite{dantzig-simplex}. The basic variable of the pivot is chosen as the smallest index such that feasibility is preserved after a pivot step. Using the dictionary oracle one can test the at most $n$ possibilities and choose the appropriate pivot. 

\section{An Output Sensitive Redundancy Detection Algorithm}
Throughout this section, we denote by $S'$ the set of nonredundant indices and by $R'$ the set of redundant indices. Denote by $LP(n,d)$ the time needed to solve an LP. By the discussion in Section \ref{sec_criss}, for any $x_r$, $r \in E$, we can find a certificate in time $LP(n,d)$.
Theorem \ref{thm_algo} presents a Clarkson type, output sensitive algorithm with running time $\mathcal{O}(d\cdot  (n+d) \cdot s^{d-1} \cdot LP(s,d) + d \cdot s^d \cdot LP(n,d))$, that for a given LP outputs the set $S'$, where $s = |S'|$. Typically $s$ and $d$ are much smaller than $n$.

\subsection{General Redundancy Detection}

\noindent
\begin{tabbing}
123 \= 123 \= 123 \= 123 \= 123 \= 123 \= 123 \= 123 \= \kill
\> Redundancy Detection Algorithm($D$,$g$,$f$);\\
\> {\bf begin }\\
\> \> $R := \emptyset, S := \emptyset$;\\
\> \> {\bf while} $R \cup S \neq E$ {\bf do}\\
\> \> \> Pick any $r \notin R \cup S$ and test if $r$ is redundant w.r.t.\ $S$; \\
\> \> \> {\bf if} $r$ redundant w.r.t.\ $S$  {\bf then} \\
\> \> \> \> $R = R \cup \{r\}$;\\
\> \> \> {\bf else} /* $r$ nonredundant w.r.t.\ $S$ */ {\bf then}\\
\> \> \> \> test if $r$ is redundant w.r.t.\ $E \setminus R$; \\
\> \> \> \> {\bf if} $r$ is nonredundant w.r.t.\ $E \setminus R$ {\bf then}\\
\> \> \> \> \> $S = S \cup \{r\}$; \\
\> \> \> \> {\bf else} /* $r$ redundant w.r.t.\ $E \setminus R$ */ {\bf then}\\
\> \> \> \> \> Find some sets $S^F \subseteq S'$ and $R^F \subseteq R'$ such that $S^F \nsubseteq S$; \\
\> \> \> \> \> $R = R \cup R^F$, $S = S \cup S^F$; \\
\> \> \> \> {\bf endif}; \\
\> \> \> {\bf endif};\\
\> \> {\bf endwhile}; \\
\> \> $S^*: = S$; \\
\> \>  output $S^*$; \\ 
\> {\bf end}.\\       
\end{tabbing}
Since in every round at least one variable is added to $S$ or $R$, the algorithm terminates. The correctness of the output can easily be verified: If in the outer loop $r$ is added to $R$, $r$ is redundant w.r.t.\ $S$ and hence redundant w.r.t.\ $S^* \supseteq S$. If in the inner loop $r$ is added to $S$, $r$ is nonredundant w.r.t.\ $E \setminus R$ and hence nonredundant w.r.t.\ $S^* \subseteq E \setminus R$.

The main issue is how to find the sets $S^F$ and $R^F$ efficiently in the last step. This will be discussed in (the proof of) Lemma \ref{lemma_inner}.

A technical problem is that we cannot test for redundancy in the dictionary oracle when $S$ does not contain a nonbasis. Therefore as long as this is the case, we fix an arbitrary nonbasis $N$ and execute the redundancy detection algorithm on $S \cup N$ instead of $S$. Since this does not change correctness or the order of the running time, we will omit this detail in the further discussion.



\begin{Theorem}
\label{thm_algo}
The redundancy detection algorithm outputs $S'$, the set of nonredundant constraints in time 
\[R(n,d,s) = \mathcal{O}  \left( \sum_{i=0}^{d-1} ((n+d) \cdot s^i \cdot LP(s,d-i) + s^{i+1} \cdot LP(n,d-i)) \right)\] and consequently in time
\[R(n,d,s) = \mathcal{O}\left(d\cdot  (n+d) \cdot s^{d-1} \cdot LP(s,d) + d \cdot s^d \cdot LP(n,d)\right).\]
\end{Theorem}
The following Lemma implies Theorem \ref{thm_algo}.

\begin{Lemma}
\label{lemma_inner}
Let $R(n,d,s)$ be the running time of the redundancy detection algorithm in $n$ basic variables, $d$ nonbasic variables and $s$ the number of nonredundant variables. Then in the last step of the inner loop some sets $S^F \subseteq S'$ and $R^F \subseteq R'$, with $S^F \nsubseteq S$, can be found in time $\mathcal{O}(R(n,d-1,s) + LP(n,d))$.
\end{Lemma}

\begin{proof}[Proof of Theorem \ref{thm_algo}]
Termination and correctness of the algorithm are discussed above.
The iteration of the outer loop of the algorithm 
takes time $\mathcal{O}(LP(s,d))$ and is executed at most $n+d$ times. By Lemma \ref{lemma_inner}, the running time of the inner loop 
is $\mathcal{O}(R(n,d-1,s) + LP(n,d))$ and since in each round at least one variable is added to $S$, it is executed at most $s$ times. Therefore the total running time is given recursively by 
\[R(n,d,s) = \mathcal{O}\left((n+d) \cdot LP(s,d) + s\cdot(R(n,d-1,s) + LP(n,d))\right).\]
The claim follows by solving the recursion and noting that $R(n,0,s)$ can be set to $\mathcal{O}(n)$.
\end{proof}

It remains to prove Lemma \ref{lemma_inner}, for which we first prove some basic results below, using the dictionary oracle setting.

\begin{Lemma}
\label{lemma_l1}
Let $D=D(B)$ be a feasible dictionary of an LP of form (\ref{LP1}) and assume $F: = \{i \in B | b_i = 0\} \neq \emptyset$. We consider the subproblem of the LP denoted $LP^F$ (with dictionary $D^F$,) that only contains the rows of $D$ indexed by $F$. Then $r \in F \cup N$ is nonredundant in LP if and only if it is nonredundant in $LP^F$.
\end{Lemma}

\begin{proof}
We only need to show the "if" part. Let $r \in F \cup N$ be nonredundant in $LP^F$ with certificate $\overline{D}^{F}$. Then there exists a sequence of pivot steps from $D^F$ to $\overline{D}^{F}$. Using the same ones on $D$ and obtaining dictionary $\overline{D}$, this is a nonredundancy certificate for $r$, since $\overline{d}_{ig} = d_{ig} > 0$ for all $i \in B \setminus F$ by the definition of $F$.
\end{proof}

\begin{Lemma}
\label{lemma_l3}
Let $D = [b, -A]$ be the dictionary of an LP of form (\ref{LP1}). Then a variable $r \in E$ is nonredundant in the LP given by $D$ if and only if it is nonredundant in the LP given by $D^{0} = [0, b, -A]$. 
\end{Lemma}

\begin{proof}
If $D(B)$ is a redundancy certificate for $r$ for some basis $B$, then $D^0(B)$ is a redundancy certificate for $r$ as well. 

For the converse, let $D = D(B)$ be a nonredundancy certificate for $r$ for some basis $B$. For simplicity assume that $B = \{1,2,\dots,n\}$. 
For now assume that $b_i> 0$ for all $i\in B$ and let $D^i$ the dictionary obtained from $D^0$ by pivoting on $b_i$, $i = 1,2,\dots,n$. We will show that at least one of the $D^i$, $i \in \{0,1,\dots,n\}$ is a nonredundancy certificate for $r$.
Since after any pivot the first column of $D^i$ stays zero, $D^i$ is a nonredundancy certificate if and only if $D^i_{.r} \leq 0$. Let $R^i = (r_1^i, r_2^i, \dots r_n^i)^T := D^i_{.r}$ for $i \geq 1$ and $R = (r_1,r_2,\dots,r_n)^T: = D^0_{.r}$. 

\textbf{Claim:} Assume that $r_i^i < 0$ for any fixed $i$ and there are at least $i-1$ additional nonpositive entries (w.l.o.g.\ we assume them to be $r^i_1, r^i_2, \dots, r^i_{i-1}$). If $R^{i}$ has a positive entry (which w.l.o.g.\ we assume to be $r^i_{i+1}$), then $r_{i+1}^{i+1} < 0$ and $r_1^{i+1}, r_2^{i+1}, \dots,  r_i^{i+1}$ are nonpositive.

If $D^0$ is not a certificate for $r$, then w.l.o.g.\ $r_1 > 0$ and hence $r_1^1 = -\frac{r_1}{b_1} <0$. Therefore by induction the lemma follows from the claim. 

Assume that $r^i_1, r^i_2, \dots, r^i_{i-1} \leq 0$, $r_i^i <0$ and $r_{i+1}^i >0$. Then we 
have $r_i>0$ and
\begin{align}
&r_{i+1}^{i} = r_{i+1} -\frac{r_{i}b_{i+1}}{b_{i}} > 0 \Leftrightarrow r_ib_{i+1} < r_{i+1}b_i \Rightarrow r_{i+1} > 0, \label{eq2} \\
&\forall j < i: r_j^i = r_j - \frac{r_ib_j}{b_i} \leq 0 \Leftrightarrow r_jb_i \leq r_ib_j. \label{eq3}
\end{align}
The following calculations show the claim.
\begin{align*}
&r_{i+1}^{i+1} = -\frac{r_{i+1}}{b_{i+1}} < 0 \Leftrightarrow r_{i+1} > 0 \text{ which holds by (\ref{eq2}}). \\
&r_{i}^{i+1} = r_i - \frac{r_{i+1} b_i}{b_{i+1}} \leq 0 \Leftrightarrow r_ib_{i+1} \leq r_{i+1}b_i \text{ which holds by (\ref{eq2}}). \\
&\forall j < i: r_j^{i+1} = r_j - \frac{r_{i+1}b_j}{b_{j+1}} \leq 0 \Leftrightarrow r_j b_{i+1}\leq r_{i+1}b_j, \\
&\text{and by (\ref{eq2}) and (\ref{eq3}), }r_jb_{i+1} = (r_jb_i)(r_ib_{i+1})\cdot\frac{1}{r_ib_i} \leq r_{i+1}b_j .
\end{align*}

Now suppose that $b_i =0$ for some $i$. Then by the nonredundancy certificate $r_i \leq 0$, and it is easy to see that $r_i^j  = r_i \leq 0$ for all admissible pivots on $b_j$. Hence we can use the above construction on the nonzero entries of $b$.
\end{proof}

\begin{proof}[Proof of Lemma \ref{lemma_inner}]
Suppose that during the execution of the algorithm, $r$ is nonredundant w.r.t.\ the current set $S$, and redundant w.r.t.\ $E \setminus R$, with \emph{feasible} redundancy certificate $D = [b, -A]$, which exists by Corollary \ref{cor_redundant}. If $b > 0$, then all nonbasic indices in $N$ are nonredundant by Theorem \ref{thm_nonred}. Choose $S^F = N$, $R^F =\emptyset$. It holds that $S^F \nsubseteq S$, since otherwise $r$ would be redundant w.r.t.\ $S$. The running time of the inner loop in this case is $LP(n,d)$.

Now if there exists $i \in B$ such that $b_i = 0$, define $F = \{i \in B | b_i =0\}$, $LP^F$ and $D^F$ as in Lemma \ref{lemma_l1}.
We now recursively find all redundant and nonredundant constraints in the $LP^F$ using Lemma \ref{lemma_l3} as follows. 
From $LP^F$ we construct another LP, denoted $LP^-$ with one less nonbasic variable, by deleting $D^F_{.g}$ (the column of all zeros), choosing any element $t \in N$ and setting $t=g$. 
Finding all redundancies and nonredundancies in $LP^-$ takes time $R(|F|,d-1,s)$. 
By Lemma \ref{lemma_l3} redundancies and nonredundancies are preserved for $LP^F$.
Therefore finding them in $LP^F$ takes time $R(|F|,d-1,s) + LP(n,d) \leq R(n, d-1,s) + LP(n,d)$, where the $LP(n,d)$ term is needed to check separately whether $t$ is redundant.
Choose $S^F$ as the set of nonredundant indices of $LP^F$ and $R^F$ as the set of redundant ones. By Lemma \ref{lemma_l1} $S^F \subseteq S'$ and $R^F \subseteq R'$. 
Since by Lemma \ref{lemma_l1} $r$ is redundant in $LP^F$, $S^F \nsubseteq S$, since otherwise $r$ would be redundant w.r.t.\ $S$.   
\end{proof}

\subsection{Strong Redundancy Detection}

In this section we show how under certain assumptions the running time of the redundancy algorithm can be improved. If we allow the output to also contain some \emph{weakly redundant} constraints (see definition below), it is basically the same as the running time of Clarkson's method.

A redundant variable $r$ is called \emph{strongly redundant} if for any basic feasible solution $\overline{x}$, $\overline{x}_r > 0$. In particular for any basic feasible solution, $r \in B$. If $r$ is redundant but not strongly redundant $r$ is called \emph{weakly redundant}.

As before let $S'$, (with $|S'| = s$,) be the set of nonredundant indices and let $R_s$, (with $|R_s| = r_s$,) and $R_w$, (with $|R_w| = r_w$,) be the set of strongly and weakly redundant indices respectively.
\begin{Theorem}
\label{thm_strong}
It is possible to find a set $S^* \supseteq S'$, $S^* \cap R_s = \emptyset$ in time 
$\mathcal{O}((n+d) \cdot LP(s + r_w, d) + (s + r_w) \cdot LP(n,d))$.
\end{Theorem}

The next corollary follows immediately.
\begin{Corollary}
If there are no weakly redundant constraints, the set $S'$ of nonredundant constraints can be found in time $\mathcal{O}((n+d) \cdot LP(s,d) + s \cdot LP(n,d))$.
\end{Corollary}

The theorem is proven using the following two lemmas, which can be verified with straight forward variable substitutions. 

\begin{Lemma} \cite{chvatal-LP}
\label{lemma_h2}
Let $(LP)$ of form (\ref{LP1}), where $(LP)$ is not necessarily full dimensional. W.l.o.g.\ $B=\{1,2,\dots,n\}$. For each $i \in \{1,2,\dots,n\}$ replace the nonnegativity constraint $x_i \geq 0$ by $x_i \geq -\epsilon^i$, for $\epsilon > 0$ sufficiently small. 
Denote the resulting LP by $(LP^\epsilon)$.
Let $D^{\sigma}$ be the output of the dictionary oracle for an arbitrary dictionary $D$ of $(LP)$. Then $(LP^\epsilon)$ is full dimensional. Furthermore in $D^{\sigma, \epsilon}$, 
the corresponding output for the $\epsilon$-perturbed version, all signs can be determined by $D^{\sigma}$, and $D^{\sigma, \epsilon}_{.g}$ has no zero entries. 
\end{Lemma}

\begin{Lemma} \cite{chvatal-LP}
\label{lemma_h1}
Let $(LP)$ and $(LP^\epsilon)$ be as in Lemma \ref{lemma_h2}.  Then any nonredundant constraint in $(LP)$ is nonredundant in $(LP^\epsilon)$ and any strongly redundant constraint in $(LP)$ is strongly redundant in $(LP^\epsilon)$.
\end{Lemma}

\begin{proof}[Proof of Theorem \ref{thm_strong}]
Replace the given LP by it's $\epsilon$-perturbed version as in Lemma \ref{lemma_h2} and run the redundancy removal algorithm, which is possible by the same lemma. 
By Lemma \ref{lemma_h1}, $S^* \supseteq S'$ and $S^* \cap R_s = \emptyset$. 
Since by Lemma \ref{lemma_h2}, the entries of the $g$-column of any dictionary $D^{\sigma, \epsilon}$ are strictly positive the algorithm never runs the recursive step and the running time follows. 
\end{proof}

\noindent
\textbf{Remark:} The $\epsilon$-perturbation makes every feasible LP full dimensional, therefore the full dimensionality assumption can be dropped for Theorem \ref{thm_strong}.

\subsection{Discussion}
In this paper, we presented new combinatorial 
characterizations of redundancy and nonredundancy in
linear inequality systems.  We also presented a combinatorial algorithm
for redundancy removal.

In contrast to the Clarkson algorithm our redundancy detection algorithm does not need the whole LP but only the combinatorial information of the dictionaries.  Although in general the running time is worse, assuming that we have no weak redundancies, our redundancy removal algorithm basically has the same running time as the Clarkson algorithm. Still, a natural goal is to improve the runtime of our algorithm in the general case and get it closer to that of Clarkson's method. We do have a first output-sensitive algorithm for combinatorial redundancy detection, but the exponential dependence on the dimension $d$ is prohibitive already for moderate $d$.

Our algorithm works in a more general setting of oriented matroids.  This means one 
can remove redundancies from oriented pseudo hyperplane arrangements efficiently.
Furthermore, the algorithm can be run in parallel. Yet, analyzing the performance may not be
easy because checking redundancy of two distinct variables simultaneously may lead to the discovery of
the same (non)redundant constraint.  This is an interesting subject of future research.

\bibliographystyle{plain}
\bibliography{redundancyFVC}

\appendix

\section{Examples}
\label{sec:examples}
In Section \ref{sec:certificates} we showed the existence of certificates in the dictionary oracle for both redundant and nonredundant variables. 
How many dictionaries are needed to detect all certificates? This number depends on the given set of linear inequalities. In \ref{sec_redundancy} (\ref{sec_nonredundancy}) we give an example where the number of dictionaries needed to detect all redundancies (nonredundancies) is linear in the number of redundant (nonredundant) variables. In \ref{sec_constant} we give an example where one dictionary suffices to detect all redundancies and nonredundancies. 
  
\subsection{Maximum Number of Bases to Detect all Redundancies}
\label{sec_redundancy}
Consider the following set of linear inequalities on $2n+1$ variables, with $|N| =n$.
\begin{align*}
x_{i} &= 1 - x_{n+i}, \hspace{0.5cm} \forall i = 1,2,\dots n \\
x_{2n+1} &= 1 - \sum_{i=1}^n x_{n+i} \\
x_i &\geq 0, \hspace{0.5cm} \forall i = 1,2, \dots 2n+1.
\end{align*}

We claim that for $i = 1, 2, \dots n$, there exists exactly one $i$-redundant basis and these bases are distinct. Furthermore the constraints $x_i \geq 0$ are nonredundant for $i = n+1, n+2, \dots, 2n + 1$.  Therefore we need a unique set of $n$ dictionaries to detect all $n$ redundancies. 





We prove the claim by enumerating the bases.
We will show that there are only the following four types of feasible bases.
\begin{enumerate}
\item $B = [n] \cup \{2n + 1\}$,
\item $B = ([n] \setminus \{i\}) \cup \{n+i,2n+1\}$ for $i \in [n]$,
\item $B = [n] \cup \{n+i\}$  for $i \in [n]$ and
\item $B = (n \setminus \{i\}) \cup \{n+i, n+j\}$ for $i,j \in [n]$, $i \neq j$.
\end{enumerate}
Note that by symmetry the bases of type $2$ (3 and 4 respectively) are all the same up to permutation of the variables. Below the corresponding dictionaries are given for $i = 1, j=2$. Correctness can be verified by appropriate pivot steps. 

\begin{center}
\begin{tabular}{ c|c|c c c c c c| }
\multicolumn{1}{r}{}
 &  \multicolumn{1}{c}{g}
 & \multicolumn{1}{c}{$n\!+\!1$}
  & \multicolumn{1}{c}{$n\!+\!2$}
  & \multicolumn{1}{c}{$n\!+\!3$} 
  & \multicolumn{1}{c}{$\cdots$}
  & \multicolumn{1}{c}{$2n\!-\!1$}
  & \multicolumn{1}{c}{$2n$}\\
  \cline{2-8}
$1$ & 1& -1 & 0 & 0 & $\cdots$ & 0 & 0 \\

$2$ & 1 & 0 & -1 & 0 & $\cdots$ & 0 & 0 \\

$3$ & 1 & 0 & 0 & -1 & $\cdots$ & 0 & 0 \\

\vdots & \vdots & \vdots & \vdots & \vdots & $\ddots$ & \vdots & \vdots \\

$n\!-\!1$ & 1& 0 & 0 & 0 & $\cdots$ & -1 & 0 \\

$n$ & 1& 0 & 0 & 0 & $\cdots$ & 0 & -1 \\

$2n\!+\!1$ & 1 & -1 & -1 & -1 & $\cdots$ & -1 & -1 \\
\cline{2-8}
\multicolumn{8}{c}{1. $B=[n]\cup\{2n+1\}$} \\
\end{tabular}
\end{center}

\begin{center}
\begin{tabular}{ c|c|c c c c c c| }
\multicolumn{1}{r}{}
 &  \multicolumn{1}{c}{g}
 & \multicolumn{1}{c}{$1$}
  & \multicolumn{1}{c}{$n\!+\!2$}
  & \multicolumn{1}{c}{$n\!+\!3$} 
  & \multicolumn{1}{c}{$\cdots$}
  & \multicolumn{1}{c}{$2n\!-\!1$}
  & \multicolumn{1}{c}{$2n$}\\
  \cline{2-8}
$n\!+\!1$ & 1& -1 & 0 & 0 & $\cdots$ & 0 & 0 \\

$2$ & 1 & 0 & -1 & 0 & $\cdots$ & 0 & 0 \\

$3$ & 1 & 0 & 0 & -1 & $\cdots$ & 0 & 0 \\

\vdots & \vdots & \vdots & \vdots & \vdots & $\ddots$ & \vdots & \vdots \\

$n\!-\!1$ & 1& 0 & 0 & 0 & $\cdots$ & -1 & 0 \\

$n$ & 1& 0 & 0 & 0 & $\cdots$ & 0 & -1 \\

$2n\!+\!1$ & 0 & 1 & -1 & -1 & $\cdots$ & -1 & -1 \\
\cline{2-8}
\multicolumn{8}{c}{2. $B=([n]\setminus\{1\}) \cup\{n+1,2n+1\}$} \\
\end{tabular}
\end{center}

\begin{center}
\begin{tabular}{ c|c|c c c c c c| }
\multicolumn{1}{r}{}
 &  \multicolumn{1}{c}{g}
 & \multicolumn{1}{c}{$2n\!+\!1$}
  & \multicolumn{1}{c}{$n\!+\!2$}
  & \multicolumn{1}{c}{$n\!+\!3$} 
  & \multicolumn{1}{c}{$\cdots$}
  & \multicolumn{1}{c}{$2n\!-\!1$}
  & \multicolumn{1}{c}{$2n$}\\
  \cline{2-8}
$1$ & 0 & 1 & 1 & 1 & $\cdots$ & 1 & 1 \\

$2$ & 1 & 0 & -1 & 0 & $\cdots$ & 0 & 0 \\

$3$ & 1 & 0 & 0 & -1 & $\cdots$ & 0 & 0 \\

\vdots & \vdots & \vdots & \vdots & \vdots & $\ddots$ & \vdots & \vdots \\

$n\!-\!1$ & 1& 0 & 0 & 0 & $\cdots$ & -1 & 0 \\

$n$ & 1& 0 & 0 & 0 & $\cdots$ & 0 & -1 \\

$n\!+\!1$ & 1 & -1 & -1 & -1 & $\cdots$ & -1 & -1 \\
\cline{2-8}
\multicolumn{8}{c}{3. $B=[n]\cup\{n+1\}$} \\
\end{tabular}
\end{center}

\begin{center}
\begin{tabular}{ c|c|c c c c c c| }
\multicolumn{1}{r}{}
 &  \multicolumn{1}{c}{g}
 & \multicolumn{1}{c}{$2n\!+\!1$}
  & \multicolumn{1}{c}{$1$}
  & \multicolumn{1}{c}{$n\!+\!3$} 
  & \multicolumn{1}{c}{$\cdots$}
  & \multicolumn{1}{c}{$2n\!-\!1$}
  & \multicolumn{1}{c}{$2n$}\\
  \cline{2-8}
$2$ & 1& 1 & -1 & 1 & $\cdots$ & 1 & 1 \\

$n\!+\!2$ & 1 & 0 & -1 & 0 & $\cdots$ & 0 & 0 \\

$3$ & 1 & 0 & 0 & -1 & $\cdots$ & 0 & 0 \\

\vdots & \vdots & \vdots & \vdots & \vdots & $\ddots$ & \vdots & \vdots \\

$n\!-\!1$ & 1& 0 & 0 & 0 & $\cdots$ & -1 & 0 \\

$n$ & 1& 0 & 0 & 0 & $\cdots$ & 0 & -1 \\

$n\!+\!1$ & 0 & -1 & 1 & -1 & $\cdots$ & -1 & -1 \\
\cline{2-8}
\multicolumn{8}{c}{4. $B=([n]\setminus \{1\}) \cup\{n+1,n+2\}$} \\
\end{tabular}
\end{center}

Consider any of the dictionaries of the forms above. One can check that any pivot step that preserves feasibility returns to one of the four types. 

Observe that $B=[n] \cup \{n+i\}$ is the only $i$-redundant basis. Similarly as above one can check that none of the nonfeasible bases are redundancy certificates. $B=[n]\cup \{2n+1\}$ is an $i$-nonredundant basis, for $i= n+1, n+2, \dots , 2n$ and bases of type $4$ are $(2n+1)$-nonredundant.

\subsection{Maximum Number of Bases to Detect all Nonredundancies}
\label{sec_nonredundancy}


Consider the following set of inequalities in $2n$ variables with $|N| = n$.
\begin{align*}
x_{i} &=  \sum_{j=1, j \neq i}^n x_{n+j} - x_{n+i} +1, \hspace{5mm} \forall i=1,2,\dots n\\
x_i & \geq 0, \hspace{0.5cm} \forall i=1,2,\dots,2n
\end{align*}

We claim that for all $i = 1,2,\dots ,n$ there exists a unique $i$-nonredundant basis and those bases are pairwise distinct. Therefore we need linearily many bases to detect all nonredundancies.



For nonredundancy certificates we only need to consider feasible bases. The dictionary corresponding to the given set of linear inequalities is
\begin{center}
\begin{tabular}{ c|c|c c c c c| }
\multicolumn{1}{r}{}
 &  \multicolumn{1}{c}{g}
 & \multicolumn{1}{c}{$n\!+\!1$}
  & \multicolumn{1}{c}{$n\!+\!2$}
  & \multicolumn{1}{c}{$\cdots$} 
  & \multicolumn{1}{c}{$2n\!-\!1$}
  & \multicolumn{1}{c}{$2n$}\\
\cline{2-7}
1 & 1 & -1 & 1 & $\cdots$ & 1 & 1 \\

2 & 1 & 1 & -1 & $\cdots$ & 1 & 1 \\

\vdots & \vdots & \vdots & \vdots & $\ddots$ & \vdots & \vdots \\

$n\!-\!1$ & 1 & 1 & 1 & $\cdots$ & -1 & 1 \\

$n$ & 1 & 1 & 1 & $\cdots$ & 1 & -1 \\

\cline{2-7}
\end{tabular}
\end{center}

Note that this is a nonredundancy certificate for $n+1, n+2, \dots , 2n$.
By symmetry all the pivot steps that preserve feasibility (i.e.\ on the entries $(i,n+i)$) yield the same dictionary up to permutation. After a pivot step on entry $(1,1+n)$ the updated dictionary is

\begin{center}
\begin{tabular}{ c|c|c c c c c c| }
\multicolumn{1}{r}{}
 &  \multicolumn{1}{c}{g}
 & \multicolumn{1}{c}{1}
  & \multicolumn{1}{c}{$n\!+\!2$}
  & \multicolumn{1}{c}{$n\!+\!3$}
  & \multicolumn{1}{c}{$\cdots$} 
  & \multicolumn{1}{c}{$2n\!-\!1$}
  & \multicolumn{1}{c}{$2n$}\\
\cline{2-8}
$n\!+\!1$ & 1 & -1 & 1 & 1 & $\cdots$ & 1 & 1 \\

2 & 2 & -1 & 0 & 2 & $\cdots$ & 2 & 2 \\

3 & 2 & -1 & 2 & 0 & $\cdots$ & 2 & 2 \\

\vdots & \vdots & \vdots & \vdots & \vdots & $\ddots$ & \vdots & \vdots \\

$n\!-\!1$ & 2 & -1 & 2 & 2 &  $\cdots$ & 0 & 2 \\

$n$ & 2 & -1 & 2 & 2 & $\cdots$ & 2 & 0 \\

\cline{2-8}

\end{tabular}
\end{center}

This basis $1$-nonredundant and the only pivot step that preserves feasibility returns to the original basis.

\subsection{Minimum Number of Bases to Detect All Redundancies and Nonredundancies}
\label{sec_constant}
It is not hard to find an example, where a single dictionary is a certificate for all variables.
By the nature of the certificates this can only happen if all basic variables are redundant and all nonbasic ones are nonredundant e.g.\ for $\boldsymbol{I}$ the all-one matrix
\begin{align*}
x_{B} &= 1 + \boldsymbol{I}x_{N} \\
x_i &\geq 0, \hspace{0.5cm} \forall i \in E.
\end{align*}






\end{document}